# Unique excitonic effects in tungsten disulphide monolayers on two-layer graphene

*Cristina E. Giusca[*,1], Ivan Rungger[1], Vishal Panchal[1], Christos Melios[1], Zhong Lin[2,3], Yu-Chuan Lin[3,4], Ethan Kahn[3,4], Ana Laura Elías[2,3], Joshua A. Robinson[3,4], Mauricio Terrones[2,3,4,5], Olga Kazakova[1]*

[1]National Physical Laboratory, Hampton Road, Teddington, TW11 0LW, United Kingdom

[2]Department of Physics, The Pennsylvania State University, University Park, PA 16802, USA

[3]Center for 2-Dimensional and Layered Materials, and Center for Atomically Thin Multifunctional Coatings (ATOMIC), The Pennsylvania State University, University Park, PA 16802, USA

[4]Department of Materials Sciences and Engineering, The Pennsylvania State University, University Park, PA 16802, USA

[5]Department of Chemistry, The Pennsylvania State University, University Park, PA 16802, USA

[*] E-mail: cristina.giusca@npl.co.uk

ABSTRACT:

Light emission in atomically thin heterostructures is known to depend on the type of materials, number and stacking sequence of the constituent layers. Here we show that the thickness of a two-dimensional substrate can be crucial in modulating the light emission. We study the layer-dependent charge transfer in vertical heterostructures built from monolayer tungsten disulphide (WS$_2$) on one- and two-layer epitaxial graphene, unravelling the effect that the interlayer electronic coupling has on the excitonic properties of such heterostructures. We bring evidence that the excitonic properties of WS$_2$ can be effectively tuned by the number of supporting graphene layers. Integrating WS$_2$ monolayers with two-layer graphene leads to a significant enhancement of the photoluminescence response, up to one order of magnitude higher compared to WS$_2$ supported on one-layer graphene. Our findings highlight the importance of substrate engineering when constructing atomically thin layered heterostructures.

The recently isolated monolayers of two-dimensional (2D) transition metal dichalcogenides provide an attractive material platform for exploring the fundamental physics of atomically thin semiconductors [1, 2, 3], as well as exploiting their potential for applications in optoelectronics [4, 5, 6], photonics [7] and fast, low-power electronics [8, 9]. Heterogeneous integration of 2D materials in vertically stacked monolayers enables the engineering of novel architectures, which offer a rich combination of optical and electronic properties not available in the individual constituents alone or their bulk counterparts. Interfacing transition-metal dichalcogenides (TMDs) with graphene has proved to be an extremely versatile approach for creating an impressive range of electronic and optoelectronic devices with varying complexity and functionality, such as field effect transistors [22], photo-responsive memory devices [23] and ultrafast, high-gain photodetectors [6, 24, 29]. For example, flexible photovoltaic devices with an external quantum efficiency of 30% [10] as well as efficient light-emitting diodes with improved performance in terms of brightness, luminescence and reduced contact resistance [11] have been demonstrated for graphene/WS$_2$/graphene heterostructures. The electronic and optical properties of TMDs/graphene heterostructures are strongly dependent on the underlying substrate [12] and the graphene synthesis technique, with heterostructures reported on mechanically exfoliated graphene [10, 11], graphene prepared by chemical vapour deposition (CVD) [17, 28] or epitaxial graphene grown on SiC [13, 21]. Compared to exfoliated or CVD synthesised graphene, epitaxial graphene on SiC holds a practical route for many applications (e.g., optoelectronics, quantum Hall effect metrology, and high speed electronics), given its compatibility with wafer-scale processing techniques, high carrier mobility, thickness uniformity and high quality over large area, without the need for post-growth transfer. Although epitaxial graphene has proved to be an ideal platform for the CVD growth of several TMDs heterostructures with chemically pristine interfaces and high quality structural properties [12, 13, 21], no attention is being paid to how the thickness of graphene affects the electronic and optical properties of heterostructures. Furthermore, the effect that the charge transfer across the interfaces has on the excitonic properties of heterostructures with varying number of graphene layers is yet to be elucidated. Previous synthetic approaches have mostly focused on optimizing the thickness, orientation and lateral size of the TMD layer on epitaxial monolayer graphene, showing substantial photoluminescence quenching [19, 21, 28, 29] and significant stiffening phonon modes due to strong interlayer coupling in these systems [19]. The different electronic structure of one-, two-, and thicker graphene layers suggests that the optoelectronic properties of graphene-based heterostructures can be affected by the graphene layer thickness, and we demonstrate this by studying the electronic and optical properties of heterostructures combining $WS_2$ with one-layer (1LG) and two-layer graphene (2LG) on SiC. We show that the behaviour of photoexcited carriers changes drastically in monolayer $WS_2$ supported on 2LG (2LG/1L$WS_2$) compared to $WS_2$ on 1LG (1LG/1L$WS_2$) giving rise to novel excitonic effects, which are a direct result of interlayer charge transfer and associated work function sensitivity to the thickness of graphene layers. The excitonic effects are evidenced by room temperature photoluminescence (PL) spectroscopy, directly correlated with scanning Kelvin probe microscopy (SKPM) and *ab-initio* calculations that evaluate the layer-dependent changes of charge transfer in $WS_2$-graphene heterostructures.

Epitaxial graphene grown by Si sublimation on semi-insulating 6*H*-SiC (0001) has been used as a platform for $WS_2$ growth. $WS_2$ layers have been subsequently synthesised via a vapour transport method involving sulfurization of tungsten trioxide at 900°C [29]. Further details on the growth are presented in the Methods section. The morphology of $WS_2$/graphene heterostructures investigated by scanning probe microscopy reveals individual islands of $WS_2$, which are predominantly monolayer in thickness, with sporadic appearance of two- or more layer domains, mostly of triangular shape, with lateral sizes typically larger than 10 μm (Figure 1).

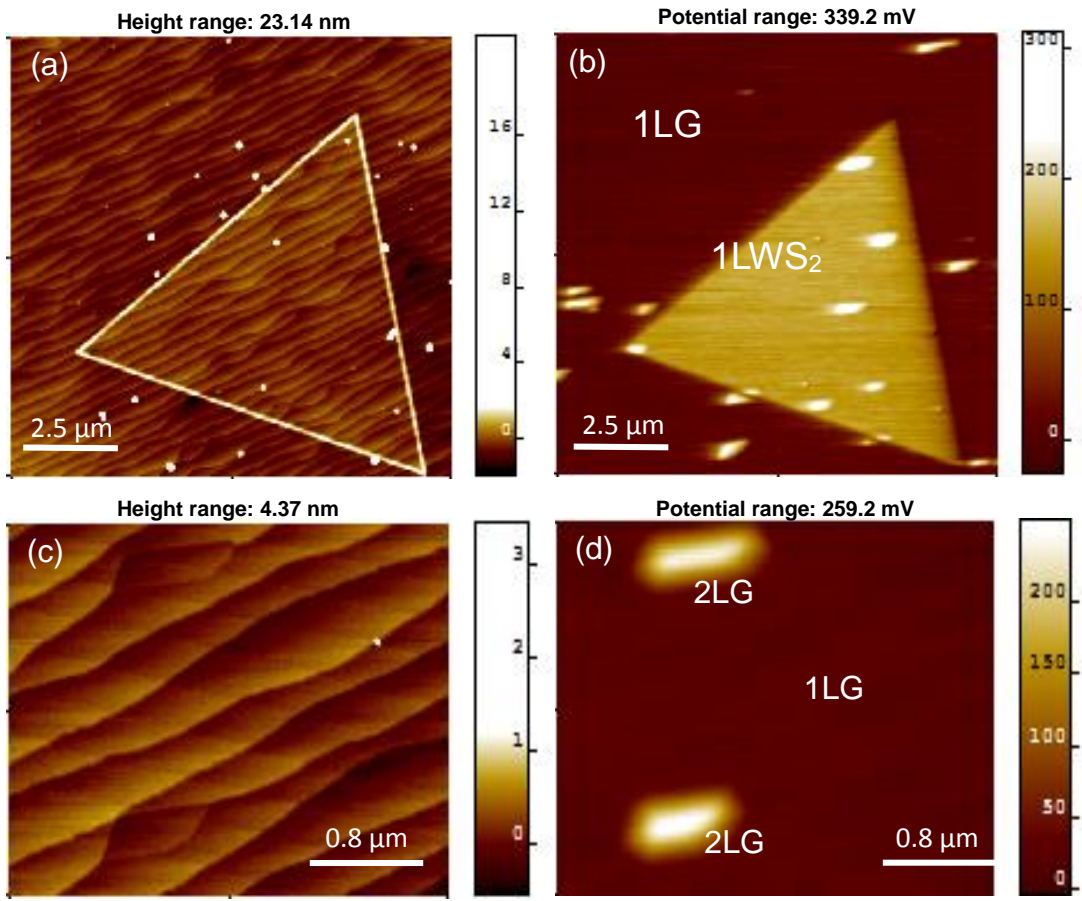

**Figure 1**: Representative (a) topography and (b) associated surface potential images of single crystal $WS_2$ grown on epitaxial graphene on SiC. The triangular $WS_2$ island has a uniform height of 0.7 nm, indicative of monolayer thickness. The morphology and surface potential recorded on a neighbouring bare graphene region are presented in (c) and (d), respectively. The dark background corresponds to 1LG and the bright patches to 2LG areas, as indicated in (d).

The topography image acquired on a bare graphene region displayed in Figure 1c reveals a terraced surface with parallel edges consistent with typical SiC morphology. While topography does not allow for a clear identification of the number of graphene layers, to date SKPM has been successfully applied to quantitatively characterise and to assess the local graphene layer thickness, see e.g. Refs. [25, 26, 27, 36]. The surface potential map of bare graphene probed by SKPM shows regions with two main distinct contrast levels: a bright one, given by small patches formed at the edge of the terraces, associated with 2LG, superimposed on a dark contrast background of one-layer graphene (1LG) (Figure 1d). The 2LG patches appear as elongated features ranging from 300 nm to ~1 μm in length and ~300 nm wide, showing a contact potential difference of 200 mV with respect to 1LG.

Remarkably, although 2LG patches are topographically located under $WS_2$ islands, their presence becomes clearly apparent on the surface potential images and is observed both on $WS_2$, as well as on the 1LG regions (Figure 1b). Apart from the 2LG patches, the distribution of surface potential across the $WS_2$ monolayer ($1LWS_2$) is relatively homogeneous, showing a higher surface potential with respect to 1LG, with a contact potential difference (CPD) of 130 mV that increases with the number of $WS_2$ layers (Figure S2f in the Supporting Information section). With respect to 2LG, 1LWS$_2$ exhibits a lower surface potential by 120 mV. Extensive previous investigations of epitaxial graphene on SiC demonstrated that graphene is *n*-doped due to charge transfer from the buffer layer, with a higher carrier (electron) density for 2LG compared with 1LG, as shown by transport measurements in our previous studies [27, 33].

To better understand the characteristics of the interaction between WS$_2$ and 1 - 2LG, as well as interpret our experimental findings, the heterostructures are investigated using density functional theory (DFT) simulations. We first address the origin of the changes in work function (*W*) for the different systems, and from it infer the experimentally found carrier concentrations in the graphene-WS$_2$ layers for the different systems. Details of the calculations are included in the Methods section.

In line with the experimental setup, we model 4 systems: 1LG, 2LG, 1LG/1LWS$_2$ and 2LG/1LWS$_2$. Figure 2a shows the side view of a 2LG/1LWS$_2$ supercell used in the DFT modelling. The value of *W* for each system is evaluated from the average of the Hartree potential over the plane of the layers ($V_H$) in the vacuum region ($V_{H,vacuum}$). If we set the Fermi energy ($E_F$) as zero, then we have $W = V_{H,vacuum}$ (Figure 2(b)). As experimentally shown in Ref. [27], *W* is strongly dependent on the substrate and environmental conditions. Therefore, we consider both the electronic charge transfer from the SiC substrate as well as adsorbates from the ambient by including effective surface charge densities placed below ($n_S$, to model the surface charges on the substrate) and above ($n_A$, to model the net charges of adsorbates) the graphene-WS$_2$ layers. Overall charge neutrality of the system imposes a resulting net charge in our DFT simulations in the graphene-WS$_2$ layers, $n_{GW}$, given by $n_{GW} = -n_S - n_A$, which corresponds to the number of carriers in the system (see Methods section for further details).

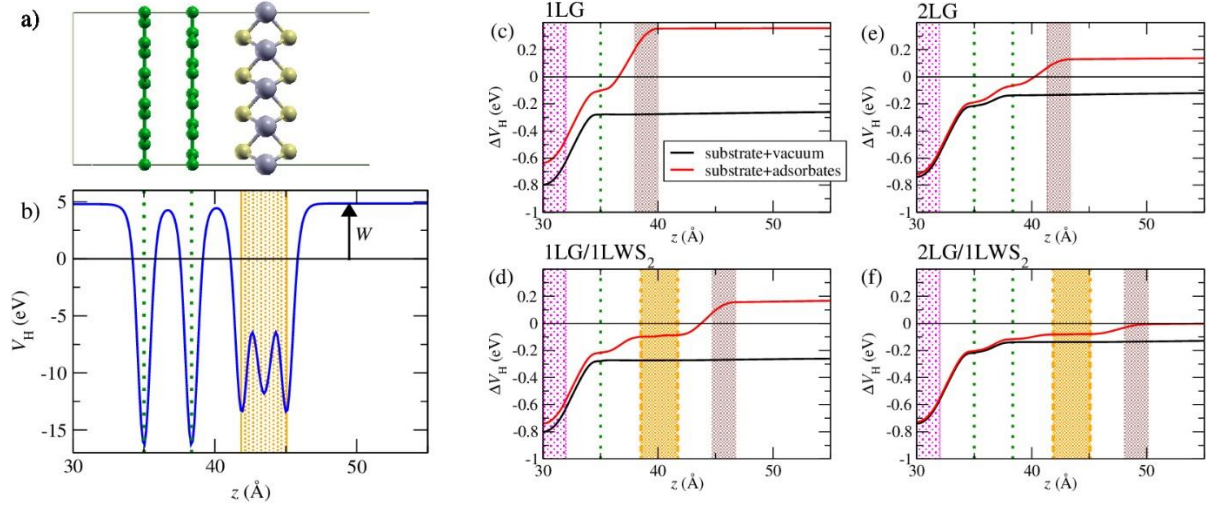

**Figure 2:** (a) Side view of a 2LG/1LWS$_2$ supercell used in the DFT modelling. (b) Planar average of the Hartree potential ($V_H$), where the green dotted vertical lines indicate the $z$ position of the graphene layers, while the yellow shaded region indicates the extension of the WS$_2$ layer. The Fermi energy ($E_F$) is set to zero, so that the value of $V_H$ in the vacuum region corresponds to the work function ($W$) of the system. (c-f) Difference, $\Delta V_H$, between $V_H$ for a system including effective substrate or adsorbate charges, and the corresponding potential for a neutral system without such added charges, with carrier concentrations given in Table 1. The vertical green dotted lines indicate the positions of the graphene layers, the orange area indicates the extension of the WS$_2$ layer, while the pink (brown) areas indicate the extension of the substrate (adsorbate) effective charge distributions.

We first evaluate the obtained difference in $V_H$ with respect to the pristine system ($\Delta V_H$) when only the substrate charges are included (black curves in Figure 2c-d). The changes in $W$ correspond to the value of $\Delta V_H$ in the vacuum region ($z > 50$ Å), and we find that this substrate induced charge transfer leads to a large reduction of W for 1LG, and to a lesser reduction for 2LG, in agreement with the SKPM data in the previous section and Ref [27]. Every charged layer leads to significant changes in $\Delta V_H$ and therefore $W$, which shows that the changes in $W$ are due to charge transfer between layers in the whole stack. The black curves are perturbed little by the $WS_2$ layer, which shows that the change of $W$ induced by the $WS_2$ layer itself is only minor, and explains why the $WS_2$ layer is largely transparent in the workfunction measurements (see Figure 1). By transparency here we mean that the experimentally measured $W$ corresponds largely to the one of the graphene substrate, modified only slightly by the presence of the $WS_2$.

Having established a calculation method to relate $W$ to the charge transfer, we can now use the experimental values of $W$ to infer the net surface charge of the adsorbates, $n_A$, when the sample is not in vacuum, as well as the resulting carrier concentration in the graphene-$WS_2$ system. The resulting values are given in Table 1 (further details on the work function calculations are given in the SI section). Importantly, it can be seen that both with and without $WS_2$ the 2LG system has a significantly higher carrier concentration than the 1LG system. Such a result was reported in Ref. [27] for the graphene-only system, and our calculations show that such an increase of carrier density for 2LG is also found for systems covered with a layer of transition metal dichalcogenide. The resulting potential profile in Fig. 2c-f with substrate and adsorbates charges (red curves) shows that the electric field is largely screened inside the $WS_2$ layer, as can be seen by an approximately horizontal $\Delta V_H$ in the $WS_2$ in Fig. 2, which reduces the total increase of $\Delta V_H$ caused by the adsorbates. The overall change of $W$ for 1LG/$WS_2$ and 2LG/$WS_2$ shown in Figure 1 is therefore due to both the screening effect and the changed number of surface adsorbates, which generally decreases with increasing number of layers (both graphene and $WS_2$).

|  | $W_{exp}$ [eV] | $n_A$ $\times 10^{13}$ cm$^{-2}$ | $n_{GW}$ $\times 10^{13}$ cm$^{-2}$ |
|---|---|---|---|
| **1LG** | 5.10 ± 0.1 | -0.81 | -0.14 |
| **2LG** | 4.90 ± 0.1 | -0.33 | -0.62 |
| **1LG/WS$_2$** | 4.97 ± 0.1 | -0.43 | -0.52 |
| **2LG/WS$_2$** | 4.85 ± 0.1 | -0.12 | -0.83 |

**Table 1:** Experimental work functions ($W_{exp}$) for the four different systems considered here, calculated effective adsorbates surface charge concentration ($n_A$) corresponding to such values, and resulting number of carriers in graphene and graphene-WS$_2$ system ($n_{GW}$) as a result of surface adsorbates. $W_{exp}$ has been obtained as described in the Methods section. The errors quoted for $W_{exp}$ originate from the tip work function determination (details in Methods), note that the $U_{CPD}$ value, which is relevant for measured differences of $W_{exp}$ between the systems, bears a smaller error of ±0.02 eV. The substrate effective surface charge density, $n_S$, is set to -0.95x10$^{13}$ cm$^{-2}$ [27].

The band structures for the neutral free-standing system, as well as for the system including the SiC substrate and adsorbates as found in experiment, with the carrier concentrations given in Table 1, are shown in Figure 3. The WS$_2$ states can be seen as additional bands below -1 eV and above +0.8 eV. While in the pristine systems (black curves) the graphene cone is always at $E_F$, for the charged systems it shifts down in energy in accordance to the carrier concentration values for the graphene-WS$_2$ system. We note that the charge-transfer induced electric field from adsorbates and SiC substrate leads to a small bandgap opening in the systems with 2 layers of graphene.

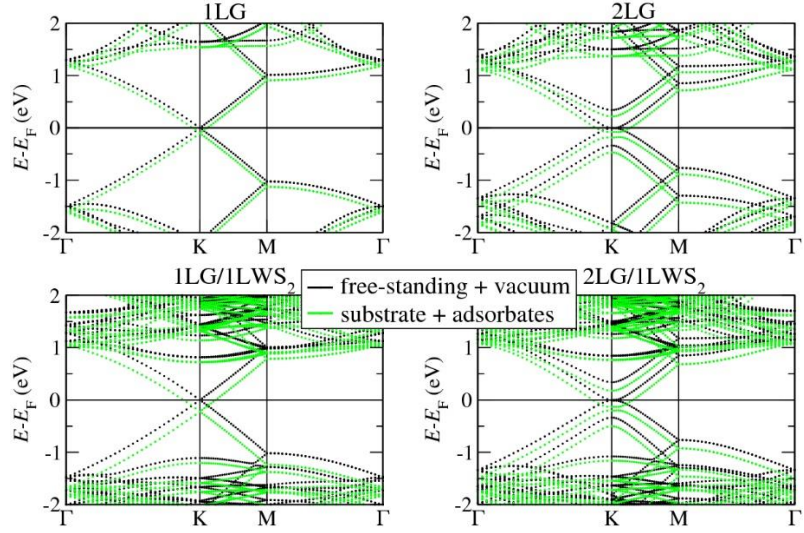

**Figure 3**: Band structures for 4 different systems: a neutral, free-standing setup with vacuum (black curves), as well as for the setup including substrate and adsorbate effective charges (green curves), with carrier concentrations given in Table 1.

Having established the charge transfer across the layers, we further investigate the optical transitions in 1LWS$_2$ by photoluminescence (PL) spectroscopy mapping to determine the impact of the underlying 1 and 2LG on the excitonic properties of WS$_2$. A representative map recorded on 1LG/1LWS$_2$ showing the spatial distribution of the PL intensity is displayed in Figure 4a.

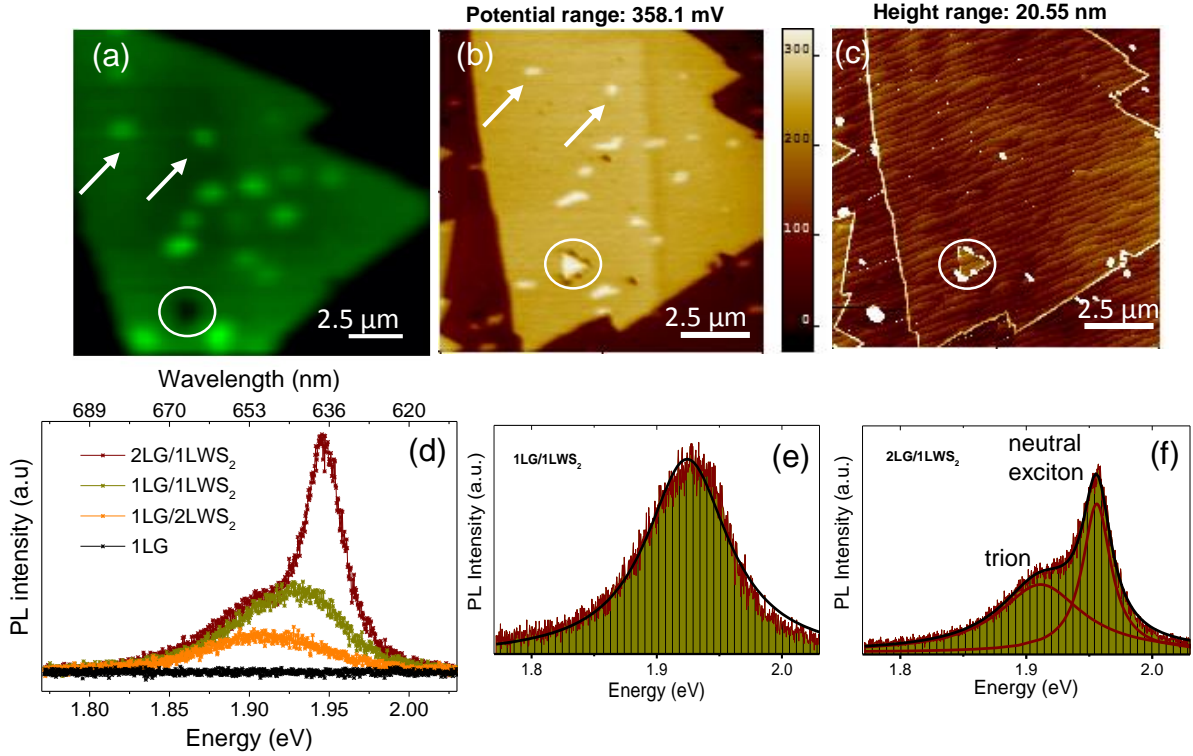

**Figure 4:** (a) Spatial distribution of PL intensity of 1LWS$_2$ grown on epitaxial graphene. Region of reduced intensity highlighted by the white circle in (a) corresponds to 2LWS$_2$ as evidenced by surface potential (b) and topography map (c). Higher intensity PL regions, as the ones indicated by arrows in (a) correspond to 2LG/1LWS$_2$, as confirmed by surface potential data in (b). 2LG patches present on the surface potential map, outside the WS$_2$ area, do not produce a PL signal. (d) Representative PL spectra for bare 1LG, 1LG/1LWS$_2$, 1LG/2LWS$_2$ and 2LG/1LWS$_2$. (e), (f) Peak deconvolution using Lorenz shape components for 1LG/1LWS$_2$ and on 2LG/1LWS$_2$, respectively. The full-width-at-half-maximum (FWHM) of the PL peak for 1LG/1LWS$_2$ is 80 meV, whereas for 2LG/1LWS$_2$, it is 80 meV for the trion and 20 meV for the neutral exciton, respectively.

The PL map shows variations in intensity across the surface at several locations (two such locations are indicated by arrows in Figure 4a and 4b), where the PL intensity appears enhanced with respect to the rest of the flake. Variations in PL peak intensity and position of 2D monolayers, such as $MoS_2$ and $WS_2$ have been observed before and attributed to several sources, such as inhomogeneous interaction with the substrate, doping by the trapped charges of the substrate, local defects in the flake and various edge configurations [30, 31]. However, these effects can be excluded in the present study as the surface potential measurements carried out on the same flake (Figure 4b) exhibit the presence of 2LG patches under $WS_2$, which show perfect correlation with the regions of increased intensity on the PL map. In addition, both surface potential and associated topography map (Figure 4c) show the presence of a small, triangular $2LWS_2$ region, which is correlated with a decreased PL signal relative to $1LWS_2$, and is consistent with the transition from a direct to an indirect band gap semiconductor [2, 30]. As expected, no PL signal from 1 and 2LG is observed on the region not covered by $WS_2$ in the PL intensity map.

It has been reported that the PL of TMD monolayers on graphene or metal substrates is strongly suppressed due to a fast non-radiative recombination process, as a result of photo-generated charge carrier transfer to the metal or graphene due to strong inter-layer electronic coupling [21, 28, 29, 31]. However, it is interesting to note that the light emission of $2LG/1LWS_2$ is less suppressed than that of $1LG/1LWS_2$ as supported by our PL observations.

Representative spectra extracted from the PL intensity map are displayed in Figure 4d. As expected, no PL signal is observed from the underlying 1LG. On 1LG, monolayer $WS_2$ exhibits a broad emission peak (the A exciton peak) at ~1.92 eV (Figure 4e), associated with the direct gap transition at the K point in the Brillouin zone [30]. Compared to $1LG/1LWS_2$, the PL intensity decreases for $1LG/2LWS_2$, consistent with the transition from a direct to an indirect band gap semiconductor, and the peak position appears red-shifted by 20 meV. In contrast to 1LG/1LWS$_2$, for 2LG/1LWS$_2$ the PL intensity is significantly enhanced and the spectral shape modified such as the PL peak is defined by a superposition of two individual resonances. As illustrated in Figure 4f, the two constituents evidenced by the PL peak deconvolution, carried out using Lorentzian shape components, are associated with the emission from neutral exciton states at 1.95 eV and from charged exciton states (trions) at 1.90 eV, respectively. The difference between the two values results in a binding energy for the trions of 50 meV. Values as high as 20-40 meV have been observed previously in WS$_2$ [16, 32] and of 20 and 30 meV have been reported in the case of monolayer MoS$_2$, MoSe$_2$, respectively [14, 15].

Since trions are created through the association of a neutral exciton with a free electron or hole, it is expected that the different carrier concentrations of 1 and 2LG can be associated with spatially inhomogeneous doping of WS$_2$ by the underlying 1 and 2LG, which will therefore influence the relative population of neutral and charged excitons in the WS$_2$ monolayer. These observations are consistent with previous reports, where the trion recombination has been evidenced on the basis of doping either via chemical methods [14] or electric-doping using a back gate [15, 32] to switch between neutral exciton PL and trion PL depending on the carrier density of the dichalcogenide monolayer. It is interesting to note, that on a cleaved graphite surface, monolayer WS$_2$ exhibits a uniform, single excitonic photoluminescence peak, without charged or bound excitons at room or low temperature [18].

DFT calculations detailed in the previous section and our previous transport measurements on similar graphene samples [27, 33] evidence that 1 and 2LG have different carrier densities, with 2LG consistently showing a higher carrier density compared to 1LG. A higher carrier density, and therefore a rich electron environment, is expected to correspondingly alter the exciton absorption of WS$_2$ and is consistent with the presence of trions associated with 2LG. Time-resolved ultrafast pump-probe spectroscopy showed that carriers in monolayer graphene indeed cause a change in the exciton resonance of monolayer WS$_2$ resting on graphene due to effective screening of the electric field between electrons and holes in WS$_2$ [20]. Therefore, photoexcited electrons and holes from WS$_2$ can efficiently transfer to 1LG and combine non-radiatively near the Dirac point, giving rise to the quenched emission we observe. Moreover, most photocarriers injected in WS$_2$ transfer to graphene in ~1 ps timescale, with near-unity efficiency, instead of recombining in WS$_2$ [20]. Our PL observations therefore suggest that the recombination in 2LG is less efficient compared to 1LG. If we consider the band structures of 1 and 2LG, one major difference is the presence of a small band gap in 2LG (see Figure 3b) due to a small charge asymmetry between the two layers [35], which can lead to a longer photocarrier decay time in this system. Indeed, time- and angle-resolved photoemission spectroscopy measurements on epitaxial graphene on SiC (0001) confirm that photoexcited carriers last longer in the excited state in 2LG compared to the lifetime of carriers in 1LG [34]. The excited charge carriers in the innermost conduction band of 2LG decay through fast, phonon-assisted interband transitions, whereas the lifetime is increased for charge carriers at the minimum of the outermost conduction where they remain trapped due to the band gap [34]. It suggests that in 1LG/1LWS$_2$ photoexcited carriers cascade down continuously and recombine, quenching the photoluminescence, whereas for 2LG/1LWS$_2$, the photoluminescence is less suppressed due to the small band gap in 2LG and consequently longer decay time, so relaxation has to rely more on radiative emission in this case.

In conclusion, our results show that the excitonic effects in WS$_2$-graphene heterostructures can be successfully tuned by the number of supporting graphene layers. As evidenced by combined SKPM measurements and *ab-initio* calculations, the work function of monolayer WS$_2$ appears transparent to the underlying 2LG and the carrier concentration is significantly larger for 2LG/1LWS$_2$ when compared to 1LG/1LWS$_2$. Whilst the photoluminescence is quenched for 1LG/1LWS$_2$, we show that the excitonic properties change significantly when an additional graphene layer is present, with enhanced photoluminescence for 2LG/1LWS$_2$. The presence of trions is evidenced in the 2LG/1LWS$_2$ heterostructures owing to a larger carrier density in 2LG compared to 1LG, as well as a longer photocarrier decay time due to the small band gap in 2LG. This type of heterostructure offers viable prospects towards exploitation in photonic applications, such as photovoltaics and photodetectors, particularly useful where a long lifetime of photoexcited carriers is expected for harvesting or storing energy devices.

METHODS

**Synthesis of WS$_2$ on epitaxial graphene** by ambient pressure chemical vapour deposition: A 1 cm x 1cm epitaxial graphene substrate with 10 mg of WO$_3$ powders placed on top was loaded into the centre of a quartz tube. The deposition was carried out at 900°C under atmospheric pressure. Sulphur vapour was generated by heating sulphur powders up to 250°C, and was carried by an Ar flow to react with the oxide powders.

**Scanning Kelvin probe microscopy** experiments were conducted in ambient, on a Bruker Icon AFM, using Bruker highly doped Si probes (PFQNE-AL) with a force constant ~ 0.9 N/m. Frequency-modulated SKPM (FM-SKPM) operated in a single pass mode has been used in all measurements. FM-SKPM operates by detecting the force gradient (dF/dz), which results in changes to the resonance frequency of the cantilever. In this technique, an AC voltage with a lower frequency ($f_{mod}$) than that of the resonant frequency ($f_0$) of the cantilever is applied to the probe, inducing a frequency shift. The feedback loop of FM-KPFM monitors the side modes, $f_0 \pm f_{mod}$, and compensates them by applying an offset DC voltage which is recorded to obtain a surface potential map. The work function for 1LG, 2LG and WS$_2$ was determined based on the respective values derived from surface potential images for the contact potential difference $\Delta U_{CPD}$ and a value for the tip work function, $\Phi_{tip}$ = (4.5±0.1) eV, according to: $e\Delta U_{CPD} = \Phi_{tip} - \Phi_{1LG, 2LG, WS2}$.

**Photoluminescence spectroscopy** maps were obtained using a Horiba Jobin-Yvon HR800 System, using a 532 nm wavelength laser (2.33 eV excitation energy) focused onto the sample through a 100x objective. Data were taken with a spectral resolution of (3.1 ± 0.4) cm$^{-1}$ and spatial resolution of (0.4 ± 0.1) μm.

**DFT calculations** are performed using the SIESTA code [37], within the local density approximation (LDA) for the exchange correlation. We use a double zeta plus polarization basis set, with cut-off radii set by an equivalent energy shift of 2.7 meV, which gives basis-orbitals with rather long cutoffs. The real space mesh is set by an equivalent cut-off energy of 2000 Ry, and we use a Fermi smearing for the occupation of the electronic states with a temperature set to 300 K. The experimental hexagonal lattice constant for WS$_2$ (graphene) is 3.15 Å [38] (1.42 Å), in good agreement with our LDA relaxed structure of 3.12 Å (1.41 Å).

For the graphene-WS$_2$ heterostructure we use a supercell that extends over 5x5 primitive unit cells of graphene, and 4x4 primitive unit cells of WS$_2$, as illustrated in Fig. 2(a) and S1 in the SI for a 2LG/1LWS$_2$ system. In the graphene-WS$_2$ supercell we fix the graphene lattice constant to its experimental value of 1.42 Å, which results in a small in-plane strain of the WS$_2$ layer of ~ 2.5% when compared to the experimental lattice, and we fix the interlayer distance between graphene layers to the experimental value of 3.354 Å. The distance between graphene and WS$_2$ calculated including van der Waals forces was taken from Ref. [39] to be 3.52 Å. In order to decouple periodic images of the slab we add a vacuum of at least 90 Å perpendicular to the plane of the layers. The coordinate system is defined in such a way that the *x-y* plane spans the plane of the layer, while the *z* direction is orthogonal to it. Since the charge transfer across the layers leads to a dipole field, we add a compensating dipole correction along *z* to eliminate the spurious electric field in the vacuum region resulting from the use of periodic boundary conditions, so that the potential in the vacuum region is constant. The charge transfer across the individual layers depends sensitively on the graphene density of states (DOS) around $E_F$, it is therefore important to use enough in-plane k-points to sample this DOS appropriately. We verified that using 16x16 k-points the potential is converged for the considered systems. In our calculations we use twice that value, namely 32x32 k-points.

Since typical molecules and layered materials bind at a distance of about 3-5 Å from the graphene, we assume the surface charges to be equally spread over such a region. While for specific molecules or substrates such distances can vary, the qualitative trends of our model will hold independently of the specific distance used. For graphene samples on SiC analogous to the ones studied here, a carrier density of $n_{GW} \approx -10^{13}$ cm$^{-2}$ was measured in vacuum [27, 33] (the negative sign is chosen to reflect the *n*-type character of graphene), so that $n_S = -n_{GW} \approx 10^{13}$ cm$^{-2}$ ($n_A \approx 0$ in vacuum). Since the value is approximately the same for 1LG and 2LG in our experiments [27], with $n_{GW} = 9.2 \times 10^{12}$ cm$^{-2}$ for 1LG, and $n_{GW} = 1.1 \times 10^{13}$ cm$^{-2}$ for 2LG, we keep it constant in all our simulations. The exact used value for simulations is n$_{GW}$= 0.95 $\times 10^{13}$ cm$^{-2}$, which corresponds to a substrate charge of $n_S$ = 0.125 e (e is the absolute value of the electron charge) per supercell.

ASSOCIATED CONTENT

**Supporting Information**

DFT calculations for the WS$_2$-graphene systems in vacuum, additional SKPM and PL data for WS$_2$-graphene. This material is available free of charge via the Internet at http://pubs.acs.org

AUTHOR INFORMATION

**Corresponding Author**

*E-mail: cristina.giusca@npl.co.uk

ACKNOWLEDGEMENTS

NPL authors acknowledge financial support from the UK's National Measurements Service under SC Graphene Project and the Graphene Flagship (No. CNECT-ICT-604391). The Penn State University authors acknowledge the financial support from the U.S. Army Research Office MURI grant W911NF-11-1-0362 and from the Penn State Center for Nanoscale Science (DMR-0820404 and DMR-1420620), as well as the National Science Foundation for the following grants: 2DARE-EFRI-1433311 and 2DARE-EFRI-1542707. Yu-Chuan Lin and Joshua Robinson acknowledge the support from the Center for Low Energy Systems Technology (LEAST), one of six centers supported by the STARnet phase of the Focus Center Research Program (FCRP), a Semiconductor Research Corporation program sponsored by MARCO and DARPA.